\documentclass[12pt]{article}

\usepackage{epsfig}

\def\Title#1{\begin{center} {\Large {\bf #1} } \end{center}}

\begin{document}

\Title{CLEO-c and CESR-c: Allowing Quark Flavor Physics to Reach its 
Full Potential.}

\bigskip\bigskip


\begin{raggedright}
{\it Ian Shipsey\index{Shipsey, I.}\\
Deprtment of Physics\\
Purdue University\\ 
West Lafayette, IN 47907, U.S.A.}
\bigskip\bigskip
\end{raggedright}

\begin{abstract}

We report on the physics potential of a proposed conversion of the
CESR machine and the CLEO detector to a charm and QCD factory:
``CLEO-c and CESR-c'' that will make crucial contributions to
quark flavor physics this decade, and may offer our best hope for
mastering non-perturbative QCD, which is essential if we are to
understand strongly coupled sectors in the new physics that lies
beyond the Standard Model.

\end{abstract}


\section{Executive Summary}

The goals of quark flavor physics are: 
to test the consistency of the Standard Model (SM)
description of quark mixing and CP violation, 
to search for evidence of new physics, and to sort between new physics 
scenarios initially uncovered at the LHC. This will require a 
range of measurements in the quark flavor changing sector of the SM 
at the per cent level. These measurements will come from a variety 
of experiments including 
BABAR and Belle and their upgrades, 
full exploitation of the facilities at Fermilab (CDF/D0/BTeV) 
and at the LHC (CMS/ATLAS/LHC-b), and experiments in rare kaon decays.

However, the window to new physics that quark flavor physics can 
provide, has a curtain drawn across it. The curtain represents hadronic 
uncertainty. The study of weak interaction 
phenomena, and the extraction of quark mixing matrix parameters remain 
limited by our capacity to deal with non-perturbative strong 
interaction dynamics.  Techniques such as lattice QCD (LQCD) directly address 
strongly coupled theories and have the potential to eventually determine 
our progress in many areas of particle physics. Recent advances in LQCD have 
produced a wide variety of calculations of non-perturbative quantities 
with accuracies in the 10-20\% level for systems involving one or two heavy
quark such as  $B$ and $D$  mesons, and $\Psi$ and $\Upsilon$
quarkonia. The techniques needed to reduce uncertainties 
to 1-2\% precision exist, but the path to
higher precision is hampered by the absence of accurate charm
data against which to test and calibrate the new theoretical techniques.

To meet this challenge the CLEO collaboration has proposed to operate 
CLEO and CESR  as a charm and QCD factory at charm threshold where the 
experimental conditions are optimal. In a three year focused program
CLEO-c will obtain charm data samples
one to two orders of magnitide larger than any previous experiment operating 
in this energy range, and with a detector that is significantly more powerful
than any previous detector to operate at charm threshold. CLEO-c has 
the potential 
to provide a unique and crucial validation of LQCD with accuracies of 1-2\%. 

If LQCD is validated, CLEO-c data will lead to a dramatic improvement 
in our knowledge of the quark couplings in 
the charm sector. In addition CLEO-c validation of lattice calculations, 
combined with 
B factory, Tevatron, and LHC data will allow a significant improvement in our 
knowledge of quark couplings in the beauty sector. The impact CLEO-c will have 
on our knowledge of the CKM matrix makes the experiment an essential step 
in the quest to understand the origin of CP violation and quark mixing. 
CLEO-c allows quark flavor physics to reach its full potential,
by enabling the heavy flavor community to draw back the curtain 
of hadronic uncertainty, and thereby
see clearly through the window to the new physics that lies beyond the SM.
Of equal importance, CLEO-c allows us to significantly advance our 
understanding and control over strongly-coupled, non-perturbatyive 
quantum field theories in general. An understanding of strongly coupled 
theories will be a crucial element in helping to interpret new
phenomena at the high energy frontier.

\begin{figure}
\begin{center}
\epsfig{figure=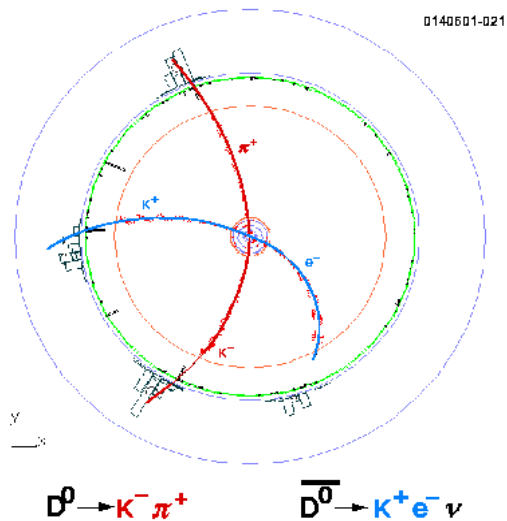,width=7cm,height=7cm}
\caption {A doubly tagged  event at the  $\psi(3770)$.}
\label{fig:cleoc_event}
\end{center}
\end{figure}

\section{Introduction}

\begin{figure}
\begin{center}
\epsfig{figure=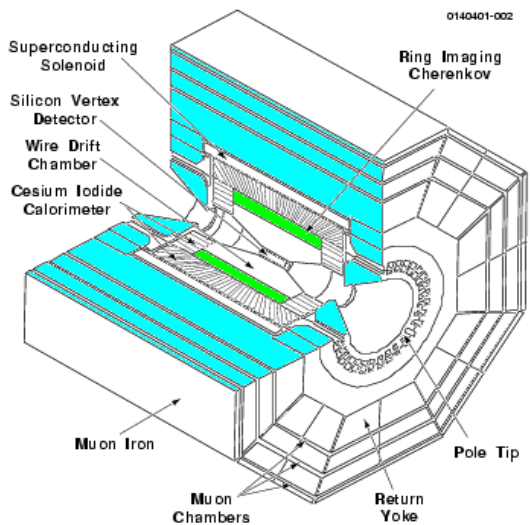,width=7cm,height=7cm}
\caption {The CLEO III detector.}
\label{fig:cleo3_det}
\end{center}
\end{figure}

For many years, the CLEO experiment at the Cornell Electron Storage
Ring, CESR, operating on the $\Upsilon$(4S) resonance, has provided most of the
world's information about the $B_{d}$ and $B_{u}$ mesons.
At the same time, CLEO, using the copious continuum pair production at the
$\Upsilon$(4S) resonance
has been a leader in the study of charm and $\tau$ physics. Now that the
asymmetric B factories have achieved high luminosity, CLEO is
uniquely positioned  to advance
the knowledge of quark flavor physics by carrying out several measurements
near charm threshold, at center of mass energies in the 3.5-5.0 GeV region.
These measurements address crucial topics which benefit from the high
luminosity and experimental constraints which exist near threshold but
have not been carried out at existing charm factories because the luminosity
has been too low, or have been carried out previously with meager statistics.
They include:

\begin{enumerate}
\item Charm Decay constants $f_D, f_{D_s}$
\item Charm Absolute Branching Fractions
\item Semileptonic decay form factors
\item Direct determination of $V_{cd}$ \& $V_{cs}$
\item QCD studies including: \\
Charmonium and bottomonium spectroscopy \\
Glueball and exotic searches \\
Measurement of R between 3 and 5 GeV, via scans \\
Measurement of R between 1 and 3 GeV, via ISR
\item Search for new physics via charm mixing, CP violation
and rare decays
\item $\tau$ decay physics
\end{enumerate}

The CLEO detector can carry out this program with only minimal
modifications. The CLEO-c project is described at length in
\cite{cleo-c} - \cite{maravin}. A very modest upgrade to the
storage ring is required to achieve the required luminosity.
Below, we summarize the advantages of running at charm threshold,
the minor modifications required to optimize the detector,
examples of key analyses, a description of the proposed run plan,
and a summary of the physics impact of the program.

\begin{table}[t]
\begin{center}
\caption{Summary of CLEO-c charm decay measurements.}
\label{tab:charm}
\begin{tabular}{cccccc}
\hline
Topic & Reaction & Energy & $L $        & current     & CLEO-c \\
      &          & (MeV)  & $(fb^{-1})$ & sensitivity & sensitivity \\
\hline
\multicolumn{1}{c}{Decay constant} &
\multicolumn{5}{c}{}\\
\hline
$f_D$ & $D^+\to\mu^+\nu$ & 3770 & 3 & UL & 2.3\% \\
\hline
$f_{D_s}$ & $D_{s}^+\to\mu^+\nu$ & 4140 & 3 & 14\% & 1.9\% \\
\hline
$f_{D_s}$ & $D_{s}^+\to\mu^+\nu$ & 4140 & 3 & 33\% & 1.6\% \\
\hline
\multicolumn{2}{c}{Absolute Branching Fractions} &
\multicolumn{4}{c}{}\\
\hline
\multicolumn{2}{c}{$Br(D^0 \to K\pi)$} & 3770 & 3 & 2.4\% & 0.6\% \\
\hline
\multicolumn{2}{c}{$Br(D^+ \to K\pi\pi)$} & 3770 & 3 & 7.2\% & 0.7\% \\
\hline
\multicolumn{2}{c}{$Br(D_s^+\to\phi\pi)$} & 4140 & 3 & 25\% & 1.9\% \\
\hline
\multicolumn{2}{c}{$Br(\Lambda_c\to pK\pi)$} & 4600 & 1 & 26\% & 4\% \\
\hline

\end{tabular}

\end{center}

\end{table}

\subsection {Advantages of running at charm threshold}

The B factories, running at the $\Upsilon$(4S) will have produced 500 million
charm pairs by 2005. However, there are significant advantages of running at
charm threshold:

\begin{enumerate}
\item Charm events produced at threshold are extremely clean.
\item Double tag events, which are key to making absolute branching fraction
measurements, are pristine.
\item Signal/Background is optimum at threshold.
\item Neutrino reconstruction is clean.
\item Quantum coherence aids $D$ mixing and CP violation studies.
\end{enumerate}

These advantages are dramatically illustrated in Figure~\ref{fig:cleoc_event},
which shows a picture of a simulated and fully reconstructed
$\psi(3770)\to D\bar{D}$ event.

\subsection {The CLEO-III Detector : Performance, Modifications and issues}

\begin{figure*}[t]
\begin{center}
\epsfig{figure=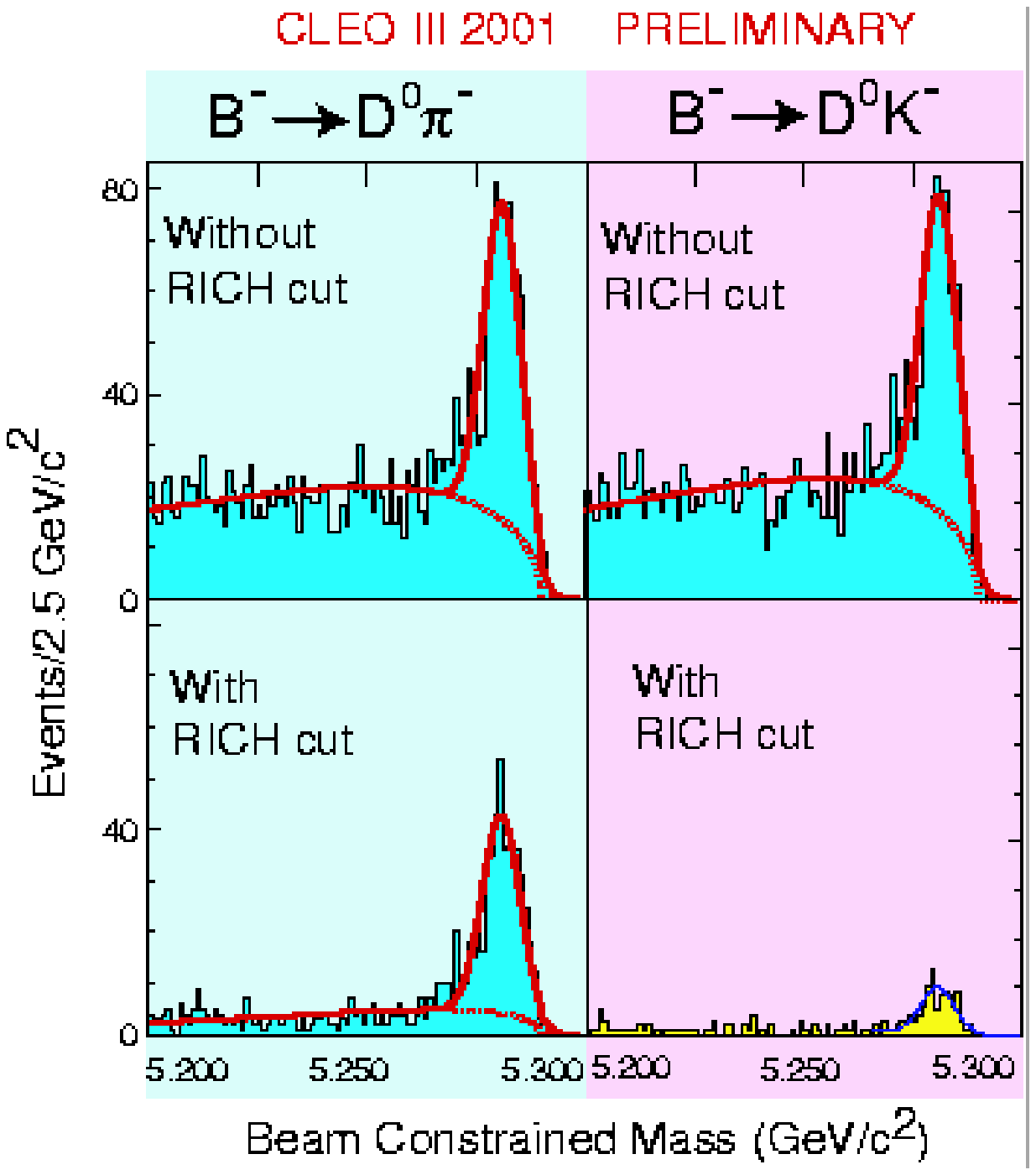,width=7.cm,height=7.cm}
\epsfig{figure=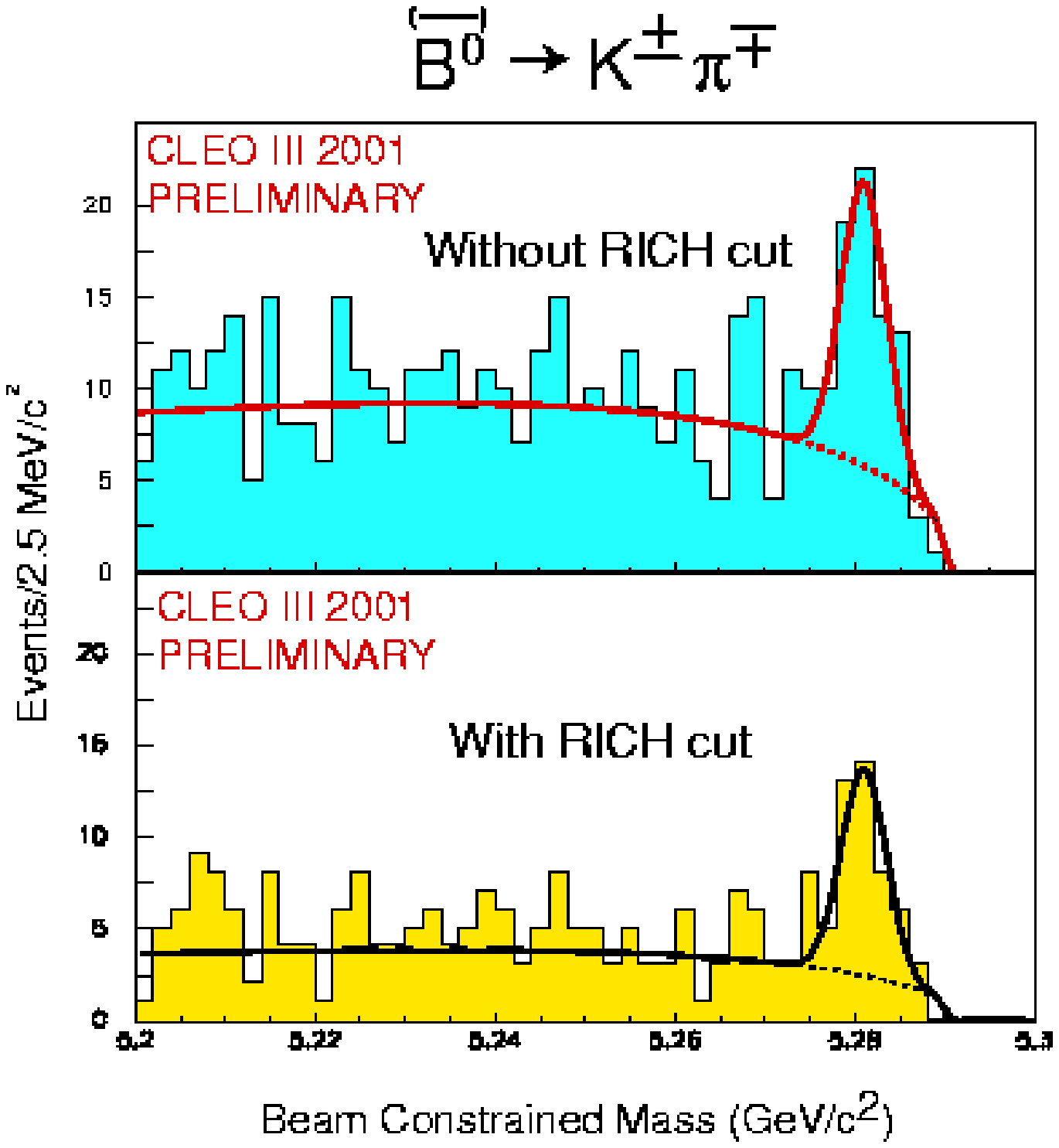,width=7.cm,height=7.cm} 
\caption
{(Left) Beam constrained mass for the Cabibbo allowed decay $B\to
D\pi$  and the Cabibbo suppressed decay $B\to DK$ with and without
RICH information. The latter decay was extremely difficult to
observe in CLEO II/II.V which did not have a RICH detector.
(Right) The penguin dominated decay $B \to K\pi$. Both of these
modes are observed in CLEO III with branching ratios consistent
with those found in CLEO II/II.V.} \label{fig:rich}
\end{center}
\end{figure*}

The CLEO III detector, shown in Figure~\ref{fig:cleo3_det}, consists of
a new silicon tracker, a new drift chamber,
and a Ring Imaging Cherenkov Counter (RICH), together  with the
CLEO II/II.V magnet, electromagnetic calorimeter and muon chambers.
The upgraded detector was installed and commissioned during the Fall of 1999
and Spring of 2000.
Subsequently operation has been very reliable (see below for a caveat)
and a very high quality data set has been obtained.
To give an idea of the power of the CLEO III detector
in Figure~\ref{fig:rich} (left plot) the beam
constrained mass for the Cabibbo allowed decay $B\to D\pi$ and
the Cabibbo suppressed decay $B\to DK$
with and without RICH information is shown.

The latter decay was extremely difficult to observe in
CLEO II/II.V which did not have a RICH detector.
In the right plot of Figure~\ref{fig:rich}
the penguin dominated decay $B \to K\pi$ is shown.
This, and other rare $B$ decay modes are observed in CLEO III 
with branching ratios consistent with those found in CLEO
II/II.V, and are also in agreement with recent Belle and BABAR results.
Figure~\ref{fig:rich} is a demonstration that CLEO III performs very well 
indeed.

Unfortunately, there is one detector subsystem that is not performing well.
The CLEO III silicon has experienced an unexpected and unexplained loss of 
efficiency. The silicon detector
will be replaced with a wire vertex chamber for CLEO-c.
We note that if one was to design a charm factory detector from
scratch the tracking would be entirely gas based to ensure
that the detector material was kept to a minimum.
CLEO-c simulations indicate that a simple six layer stereo tracker
inserted into the CLEO III
drift chamber, as a silicon detector replacement, would provide a system
with superior momentum resolution
compared to the current CLEO III tracking system.

Due to machine issues we plan to lower the solenoid field
strength to 1.0 T from 1.5 T. All other
parts of the detector do not require modification.
The dE/dx and Ring Imaging Cerenkov counters
are expected to work well over the CLEO-c momentum range.
The electromagnetic calorimeter
works well and has fewer photons to deal with at 3-5 GeV than at 10 GeV.
Triggers will work as before.
Minor upgrades may be required of the Data Acquisition system to
handle peak data transfer rates. The conclusion is that, with the addition of 
the
replacement wire chamber, CLEO is expected to work well in the 3-5 GeV
energy range at the expected rates.

\subsection {Machine Conversion}

Electron positron colliders are designed to operate optimally within 
a relatively narrow
energy range. As the energy is reduced below design, there is a 
significant reduction in synchrotron radiation, which is the primary means
of cooling the beam. In consequence, the luminosity drops, roughly as the  
beam energy to the fourth power. Without modification to the machine, 
CESR performance in the 3-5 GeV
energy range would be modest, 
well below $10^{31} {\rm cm}^{-2} {\rm s}^{-1}$.  
CESR conversion to CESR-c requires 18 m of wiggler magnets, to increase
transverse cooling,
at a cost of $\sim$ \$4M. With the wigglers installed, 
CESR-c is expected to 
achieve a luminosity in the range $2-4 \times 10^{32} {\rm cm}^{-2} 
{\rm s}^{-1}$
where the lower (higher) luminosity corresponds to $\sqrt{s} = 3.1 (4.1) 
{\rm GeV}$.

\subsection {Examples of analyses with CLEO-c}

The main targets for the CKM physics program at CLEO-c are
absolute branching ratio
measurements of hadronic, leptonic and semileptonic decays.
The first of these provides an absolute
scale for all charm and hence all beauty decays.
The second measures decay constants and the third
measures form factors and, in combination with theory, allows
the determination of $V_{cd}$ and $V_{cs}$.

\subsubsection {Absolute branching ratios}

\begin{figure}
\begin{center}
\epsfig{figure=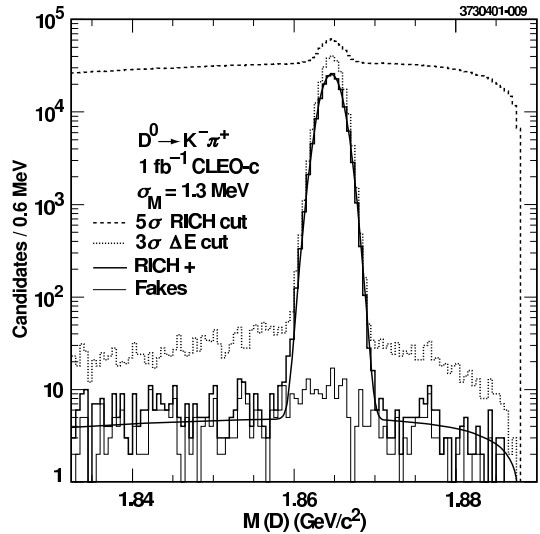,width=7cm,height=7cm}
\caption
{$K\pi$ invariant mass in $\psi(3770)\to D\bar{D}$ events
showing a strikingly clean signal for
$D\to K\pi$. The y axis is a logarithmic scale. The signal to background ratio
is $\sim$ 5000/1.}
\label{fig:d0kpi}
\end{center}
\end{figure}

\begin{figure}
\begin{center}
\epsfig{figure=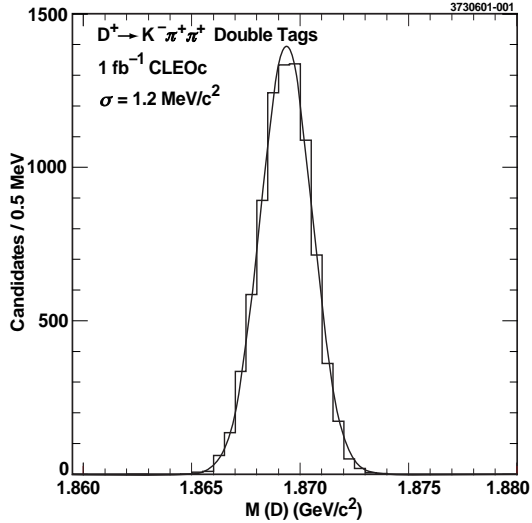,width=7cm,height=7cm}
\caption
{$K\pi\pi$ invariant mass in $\psi(3770)\to D\bar{D}$
events where the other D in the event has
already been reconstructed.
A clean signal for $D\to K\pi\pi$ is observed and the absolute
branching ratio $Br(D\to K\pi\pi)$ is measured by counting events
in the peak.}
\label{fig:br_dkpipi}
\end{center}
\end{figure}

The key idea is to reconstruct a $D$ meson in any hadronic mode.
This, then, constitutes the tag. Figure~\ref{fig:d0kpi}
shows tags in the mode $D\to K\pi$. Note the y axis is a log scale.
Tag modes are very clean. The signal to background ratio is $\sim$ 5000/1
for the example shown.
Since $\psi(3770) \to D\bar{D}$, reconstruction of a second D meson in a
tagged event to a final state X, corrected by the efficiency which is very
well known, and divided by
the number of D tags, also very well known, is a measure of
the absolute branching ratio
$Br(D\to X)$. Figure~\ref{fig:br_dkpipi} shows the $K^{-}\pi^{+}\pi^{+}$
signal from doubly tagged events. It is approximately background free.
The simplicity of $\psi(3770) \to D\bar{D}$
events combined with the absence of background allows the determination of
absolute branching ratios with extremely small systematic errors.
This is a key advantage of running at threshold.

\subsubsection {Leptonic decay $D_s\to\mu\nu$}

This is a crucial measurement because it provides information which
can be used to extract the weak decay constant, $f_{D_{s}}$. The
constraints provided by running at threshold are critical to extracting
the signal.

The analysis procedure is as follows:
\begin{enumerate}
\item Fully reconstruct one $D_s$, this is the tag.
\item Require one additional charged track and no additional photons.
\item Compute the missing mass squared ($m_{\nu}^2$) which  peaks at zero for
a decay where only a neutrino is unobserved.
\end{enumerate}

\begin{figure*}
\begin{center}
\epsfig{figure=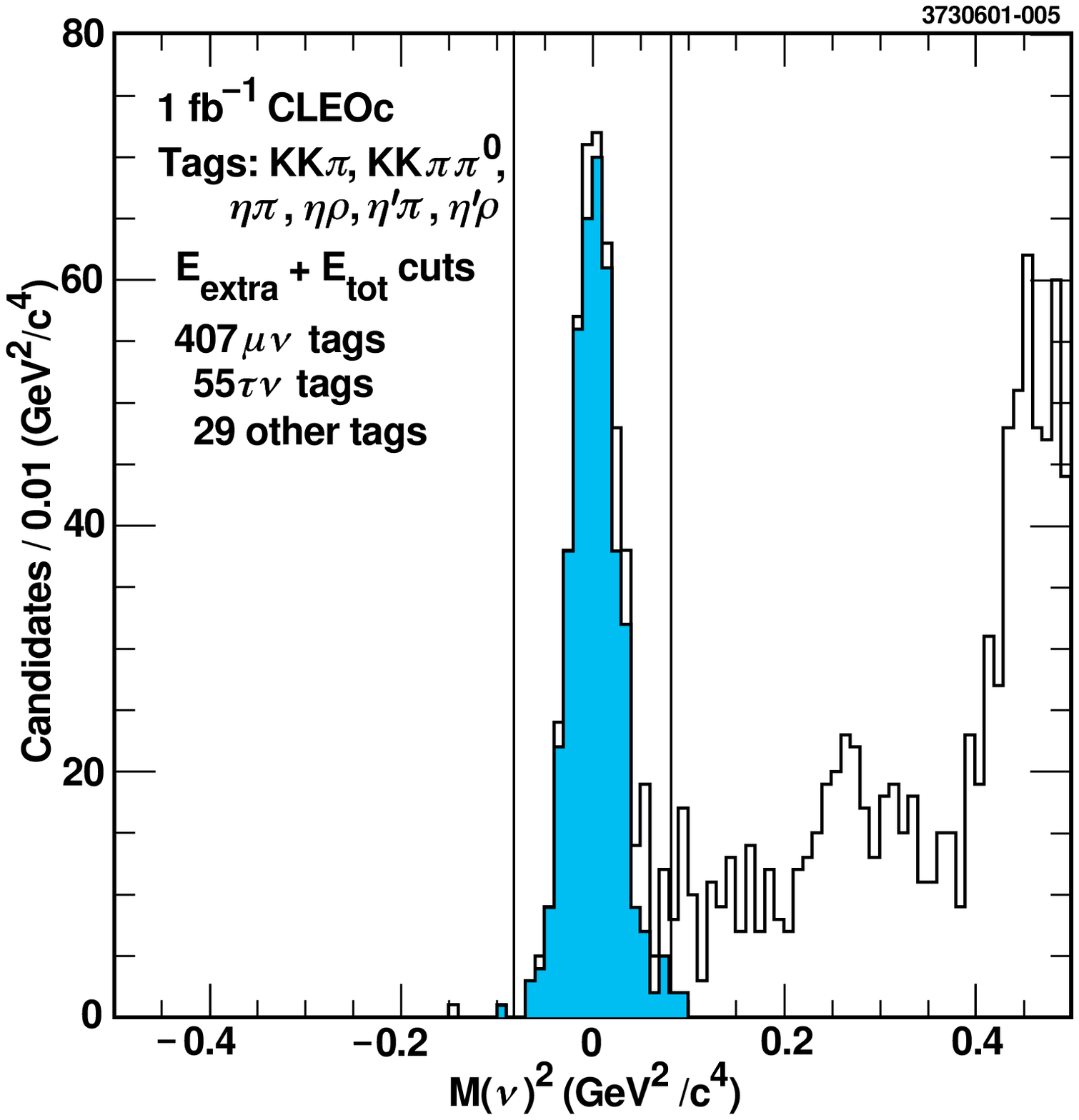,width=6cm,height=6cm}
\epsfig{figure=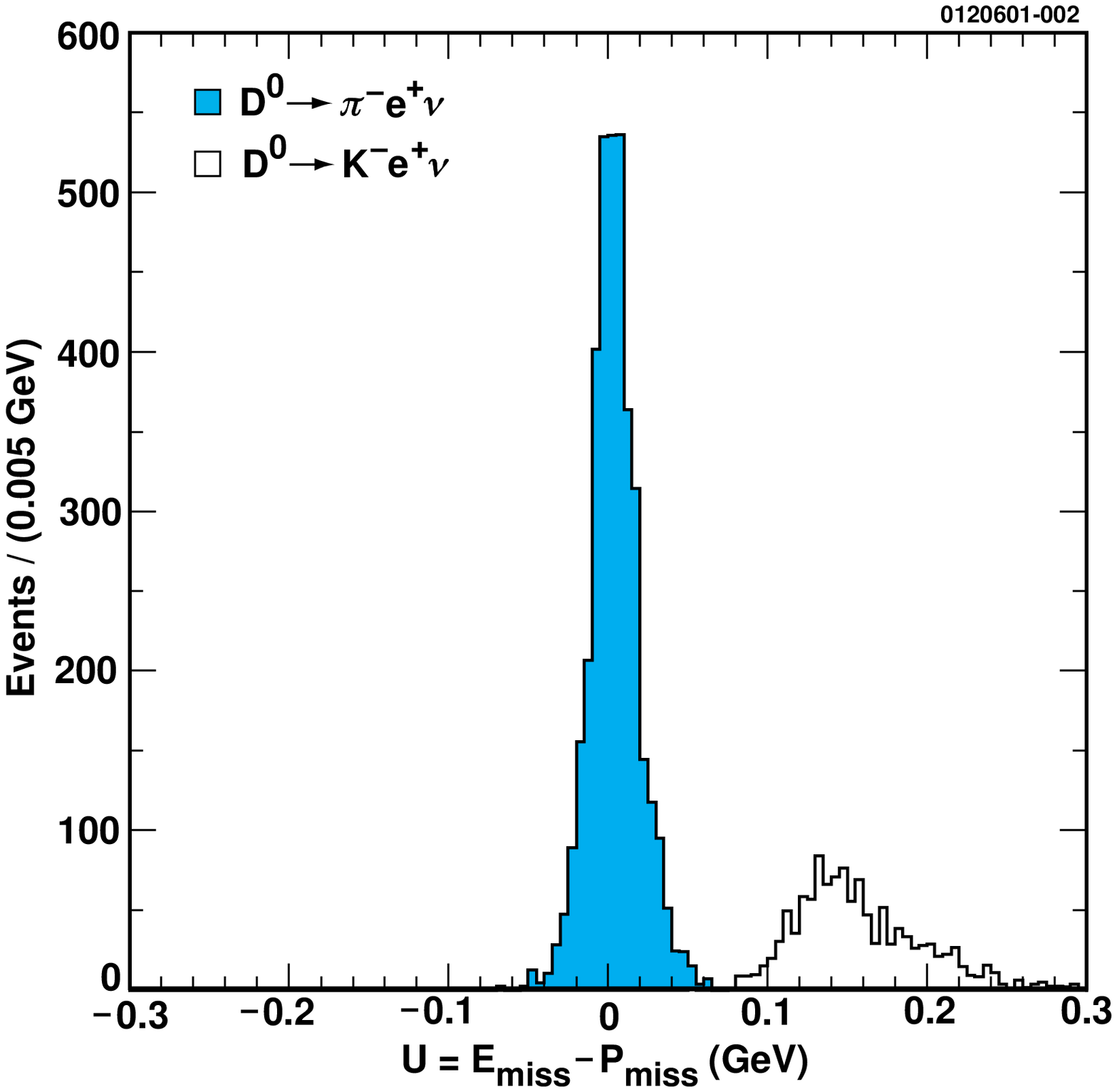,width=6cm,height=6cm}
\caption
{(Left) Missing mass squared for $D_{s} \bar{D_{s}}$ tagged pairs produced
at $\sqrt{s}=4100$ MeV. Events due to the decay $D_s\to\mu\nu$ are shaded.
(Right) The difference between the missing energy and missing
momentum in $\psi(3770)\to D\bar{D}$ tagged events for
the Cabibbo suppressed decay $D\to \pi \ell\nu$ (shaded).
The unshaded histogram arises from the ten times more copiously produced
Cabibbo allowed
transition $D\to K \ell\nu$ where the $K$ is outside the fiducial volume of
the RICH.}
\label{fig:munu_pienu}
\end{center}
\end{figure*}

The missing mass resolution, which  is of order $\sim m_{\pi^0}$,
is sufficient to reject the backgrounds to this process as shown
in Fig.~\ref{fig:munu_pienu}. There is no need to identify muons,
which helps reduce the systematic error. One can inspect the
single prong to make sure it is not an electron. This provides a
check of the background level since the leptonic decay to an
electron is severely helicity-suppressed and no signal is expected
in this mode.

\subsubsection {Semileptonic decay $D\to \pi e^+\nu$}

The analysis procedure is as follows:
\begin{enumerate}
\item Fully reconstruct one D, this constitutes the tag.
\item Identify one electron and one hadronic track.
\item Calculate the variable, $U = E_{miss} - P_{miss}$, which
peaks at zero when only a neutrino has escaped detection, which is the case
for semileptonic decays.
\end{enumerate}
Using the above procedure results in the right plot of
Figure~\ref{fig:munu_pienu}.
With CLEO-c for the first time it will become possible to make
precise branching ratio and
absolute form factor measurements of every charm meson
semileptonic pseudoscalar to
pseudoscalar  and pseudoscalar to vector transition.
This will be a lattice validation data set without
equal. Figure~\ref{fig:error_br} shows the current precision
with which the absolute semileptonic branching ratios
of charm particles are known, and the precision attainable
with CLEO-c.

\subsection {Run Plan}

CLEO-c must run at various center of mass energies to
achieve its physics goals. The ``run plan'' currently used
to calculate the physics reach is given below. This plan assumes CESR-c 
achieves design luminosity.  Item 1 is prior
to machine conversion, while the remaining items are post machine conversion.

\begin{enumerate}
\item 2002 : $\Upsilon$'s --  1-2 $fb^{-1}$ each at
$\Upsilon(1S),\Upsilon(2S),\Upsilon(3S)$ \\
Spectroscopy, electromagnetic transition matrix elements, the leptonic width.
$\Gamma_{ee}$, and searches for the yet to be discovered
$h_b, \eta_b$ with 10-20 times the existing world's data sample. 
As of July 2002,
most of this data has been collected. 
\item 2003 : $\psi(3770)$ -- 3 $fb^{-1}$ \\
30 million events, 6 million tagged D decays (310 times MARK III)
\item 2004 : 4100 MeV -- 3 $fb^{-1}$ \\
1.5 million $D_{s}D_{s}$ events, 0.3 million tagged $D_s$ decays
(480 times MARK III, 130 times BES)
\item 2005 : $J/\psi$ -- 1 $fb^{-1}$ \\
1 billion $J/\psi$ decays (170 times MARK III, 20 times BES II)
\end{enumerate}

\subsection{Physics Reach of CLEO-c}

\begin{figure}
\begin{center}
\epsfig{figure=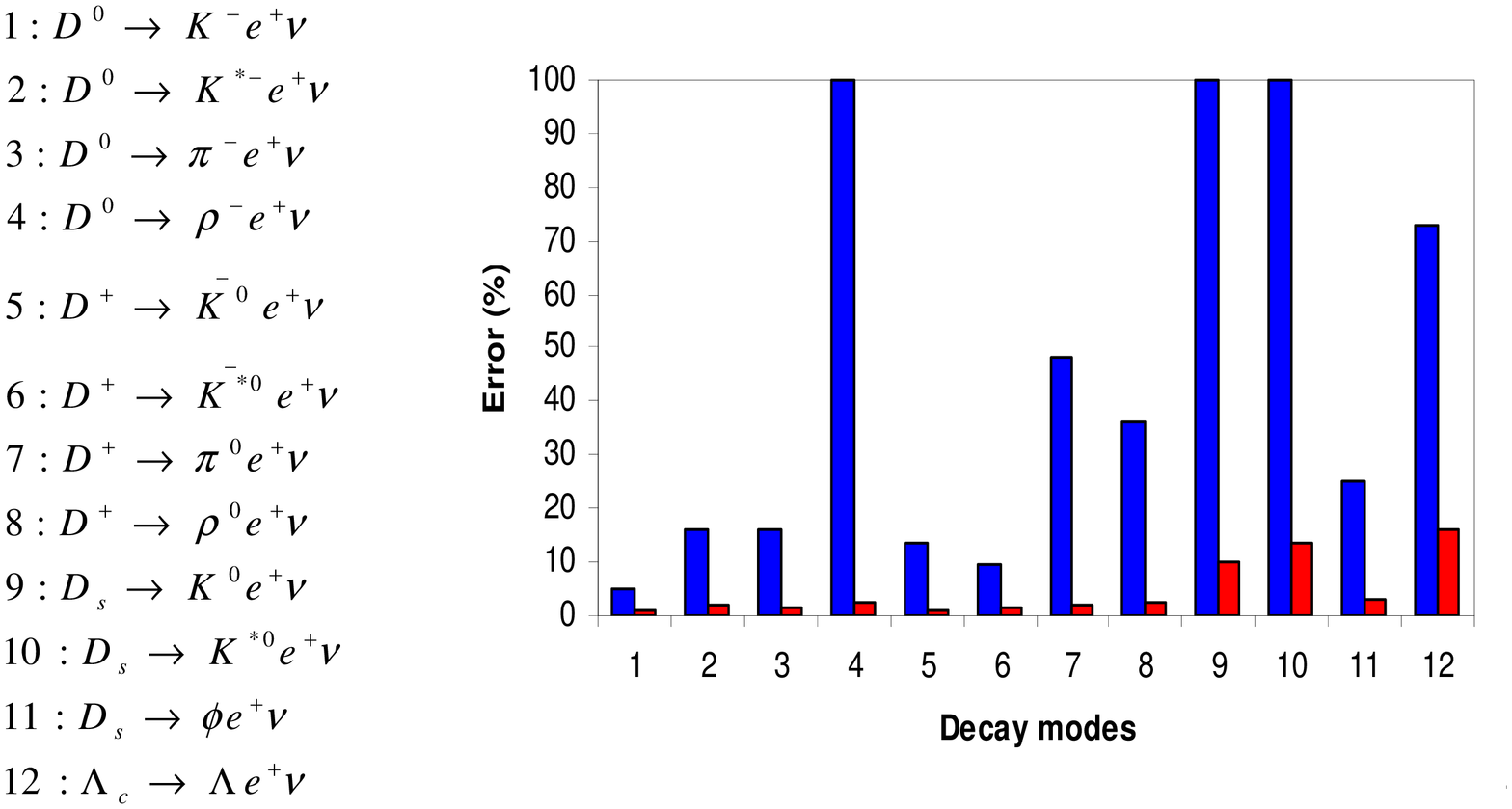,height=11cm,angle=-90}
\caption
{Absolute branching ratio current precision from the PDG
(left entry) and precision
attainable at CLEO-c (right entry ) for twelve semileptonic charm decays.}
\label{fig:error_br}
\end{center}
\end{figure}

\begin{figure}
\begin{center}
\epsfig{figure=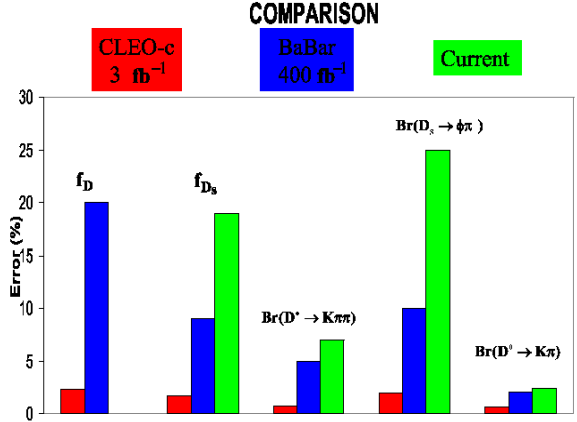,width=8cm,height=6cm}
\caption
{Comparison of CLEO-c (left) BABAR (center) and PDG2001 (right)
for the charm meson decay constants and three important charm meson
hadronic decay branching ratios.}
\label{fig:comparison}
\end{center}
\end{figure}

Tables~\ref{tab:charm}, \ref{tab:ckm} , and \ref{tab:comparison},
and Figures~\ref{fig:error_br} and
\ref{fig:comparison} summarize the CLEO-c measurements
of charm weak
decays, and compare the precision obtainable with CLEO-c
to the expected precision at BABAR
which expects to have recorded about 500 million charm pairs by 2005.
While BABAR data allows improvement in the precision with which
these quantities can be measured,  CLEO-c clearly achieves far greater
precision for many measurements.
The reason for this is the ability to measure absolute branching
ratios by tagging, and the absence of background at threshold.
For charm quantities where CLEO-c is
not dominant, it will remain comparable in sensitivity, and complementary 
in technique, to the B factories.
Also shown in Table~\ref{tab:comparison} is a summary of the data set size
for CLEO-c and BES II at the $J/\psi$ and $\psi'$,
and the precision with which R, the ratio of the $e^{+}e^{-}$ annihilation
cross section into hadrons to mu pairs, can be measured.
The CLEO-c data sets are over an order of magnitude larger, the precision
with which R is measured is a factor of three higher,
in addition the CLEO detector is vastly
superior to the BES II detector.

Taken together the CLEO-c datasets at the $J/\psi$ and $\psi'$ will be
qualitatively and quantitatively superior to any previous dataset
in the charmonium sector thereby
providing discovery potential for glueballs and exotics without equal.

\begin{table}
\begin{center}
\caption{Summary of direct CKM reach with CLEO-c}
\label{tab:ckm}
\begin{tabular}{cccccc}
\hline
Topic & Reaction & Energy & $L $        & current     & CLEO-c \\

      &          & (MeV)  & $(fb^{-1})$ & sensitivity & sensitivity \\
\hline
$V_{cs}$ & $D^0\to K\ell^+\nu$ & 3770 & 3 & 16\% & 1.6\% \\
\hline
$V_{cd}$ & $D^0\to\pi\ell^+\nu$ & 3770 & 3 & 7\% & 1.7\% \\
\hline
\end{tabular}
\end{center}
\end{table}

\begin{table}
\begin{center}
\caption{Comparision of CLEO-c reach to BABAR and BES}
\label{tab:comparison}
\begin{tabular}{cccccc}
\hline
Quantity & CLEO-c & BaBar  & Quantity & CLEO-c & BES-II \\
\hline
$f_D$ & 2.3\% & 10-20\%    & \#$J/\psi$ & $10^9$ & $5\times 10^7$ \\
\hline
$f_{D_s}$ & 1.7\% & 5-10\% & $\psi'$   & $10^8$ & $3.9\times 10^6$ \\
\hline
$Br(D^0 \to K\pi)$ & 0.7\% & 2-3\% & 4.14 GeV & $1 fb^{-1}$ &$23 pb^{-1}$\\
\hline
$Br(D^+ \to K\pi\pi)$ & 1.9\% & 3-5\%  & 3-5 R Scan & 2\%  & 6.6\% \\
\hline
$Br(D_s^+\to\phi\pi)$ & 1.3\% & 5-10\% & \multicolumn{3}{c}{} \\
\hline
\end{tabular}
\end{center}
\end{table}

\subsection{CLEO-c Physics Impact}

CLEO-c will provide crucial validation of Lattice QCD,
which  will be able to  calculate with
accuracies of 1-2\%.
The CLEO-c decay constant and semileptonic data will provide a ``golden'',
and timely test
while CLEO-c QCD and charmonium data provide additional benchmarks.
CLEO-c will provide dramatically improved  knowledge of absolute
charm branching fractions
which are now contributing significant errors to measurements involving
b's in a timely fashion.
CLEO-c will significantly improve knowledge of CKM matrix elements
which are now not very well known.
$V_{cd}$ and $V_{cs}$ will be determined directly by CLEO-c data and LQCD,
or other theoretical techniques.
$V_{cb}, V_{ub}, V_{td}$
and $V_{ts}$ will be determined with enormously improved precision
using B factory and Tevatron data, once the
CLEO-c program of lattice validation is complete.
Table~\ref{tab:ckm_summary} provides a summary of the situation.
CLEO-c data alone will also allow new tests of the unitarity of the CKM matrix.
The unitarity of the second
row of the CKM matrix will be probed at the 3\% level.
CLEO-c data will also test unitarity by measuring
the ratio of the long sides of the
squashed $cu$ triangle to 1.3\%.

Finally the potential to observe new forms of matter; glueballs,
hybrids, etc. in $J/\psi$ decays and new
physics through sensitivity to charm mixing, CP violation,
and rare decays provides a discovery
component to the program.

\begin{table}
\begin{center}
\caption{Current knowledge of CKM matrix elements (row one). Knowledge of
CKM matrix elements after CLEO-c (row two).
The improvement in the precision with which 
$V_{cd}$ and $V_{cs}$ are known is attainable with
CLEO-c data combined with Lattice QCD.
The improvement in precision with which $V_{cb}$, $V_{ub}$, $V_{td}$, and  
$V_{ts}$ are known is obtained from CLEO-c validated Lattice 
QCD calculations and B factory and Tevatron
data.}
\vspace{0.1in}
\label{tab:ckm_summary}
\begin{tabular}{cccccc}
\hline
$V_{cd}$ & $V_{cs}$ & $V_{cb}$ & $V_{ub}$ & $V_{td}$ & $V_{ts}$ \\
\hline
7\% & 16\% & 5\% & 25\% & 36\% & 39\% \\
\hline
1.7\% & 1.6\% & 3\% & 5\% & 5\% & 5\% \\
\hline
\end{tabular}
\end{center}
\end{table}

\bigskip


I would like to thank my CLEO colleagues for providing the opportunity 
to represent the collaboration at this conference. It is a privilege to 
be part of the CLEO collaboration. I thank
Ikaros Bigi, Gustavo Burdman, Andreas Kronfeld, Peter Lepage, Zoltan Ligeti 
and Matthias Neubert for valuable discussions. Finally, I thank Nigel Lockyer
and his support team for the superb organization of this conference.


\end{document}